\documentclass[aps]{revtex4}
\usepackage{amsmath}
\usepackage{graphicx}
\usepackage{epsfig,amsmath}

\begin{document}

\title{Reconnections of Vortex Loops in the Superfluid Turbulent HeII. Rates of the
Breakdown and Fusion processes.}
\author{Sergey K. Nemirovskii}
\affiliation{Institute for Thermophysics, Lavrentyeva, 1, 630090
Novosibirsk, Russia}
\date{today}

\begin{abstract}
Kinetics of merging and breaking down vortex loops is the
important part of the whole vortex tangle dynamics. Another part
is the  motion of individual lines, which obeys the Biot-Savart
law in presence of friction force and of applied external velocity
fields if any. In the present work we evaluate the
coefficients of the reconnection rates $A(l_{1},l_{2},l)$ and $%
B(l,l_{1},l_{2})$. Quantity $A$ is a number (per unit of time and
per unit of volume) of  events,  when  two loops with lengths
$l_{1}$and $l_{2}$ collide and form the single loop of length $\ l=l_{1}+l_{2}$. Quantity $%
B(l,l_{1},l_{2}) $ describes the rate of events, when the single
loop of the length $l$ breaks down into two the daughter loops of
lengths $\ l_{1}$ and $l_{2}$. These quantities
ave evaluated as the averaged numbers of zeroes of vector $\mathbf{S}%
_{s}(\xi _{2},\xi _{1},t)$ connecting two points on the loops of
$\xi _{2}$ and $\ \xi _{1}$ at moment of time $t$. Statistics of
the individual loops is taken from  the Gaussian model of vortex
tangle.\\
PACS-number 67.40
\end{abstract}

\maketitle

\section{Introduction}

Quantized vortices appeared in quantum fluids and other systems play a
fundamental role in the properties of the latter. For that reason they have
been an object of intensive study for many years (for review and
bibliography see e. g. \cite{Don_book}). The greatest success in
investigations of dynamics of quantized vortices has been achieved in
relatively simple cases such as a vortex array in rotating helium or vortex
rings. However these simple cases are rather exception than a rule. Due to
extremely involved dynamics initially straight lines or rings evolve to form
highly chaotic structure, so called vortex tangle (see e.g. papers \cite
{Feynman}, \cite{Vinen} and review \cite{NF}). The vortex tangle consists of
a set of vortex loops of different lengths. The individual vortex loops
evolves obeying the Biot-Savart law, undergoing the friction force and the
applied external velocity field. All these processes are good understood,
although the according equations of motion are highly nonlinear (and
nonlocal) therefore general analysis can be made with use of numerical
methods.

\begin{figure}[tbp]
\includegraphics[width=7cm]{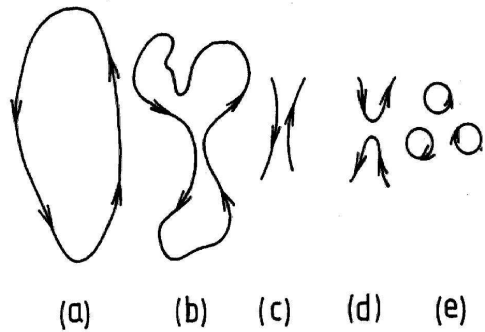}
\caption{The reconnection process schematically (Feynman, 1955, Fig.\ 10). (a)
initial stage, (b) and (c) are stages of collapse, (d) reconnection stage,
(e) stage of degeneration of vortex rings into thermal excitations.}
\label{Feynman}
\end{figure}

Besides motion of each of the individual loops there is one more
element of general evolution of the vortex tangle related to
collision of loops, or intersection of elements of vortex lines.
During intersection of lines the very complicated process, related
to arrangement of the vortex core takes place \cite{Koplik}.
Nevertheless this process is relatively short, therefore it is
usually accepted that the filaments instantly reconnect whenever
they intersect each other. It is widely appreciated that the
reconnection processes influence both the structure and evolution
of vortex tangle. However the questions how it happens and what
mechanisms are responsible for these remain open. For instance,
Feynman in his pioneering paper devoted to superfluid turbulence
proposed scenario how the vortex tangle decays in absence of
applied external counterflow. According to this scenario a fusion
of small vortex rings into larger ones as well as a breakdown into
smaller ones is possible at the moments of the reconnection events
(see Fig.~\ref{Feynman}).
In assumption that \textit{\ on the average} the last property
dominates i.e. the cascade like process of formation of smaller
and smaller loops forms. When the scale of the small rings becomes
of the order of the interatomic distances, which is the final
stage of the cascade, the vortex motion is degenerated into
thermal excitations. This key idea that degeneration of the vortex
tangle occurs due to cascade-like transferring of the length in
space of scale of sizes of vortex loops was indirectly confirmed
only in numerical calculations, where the procedure of artificial
elimination of small loops had been used
\cite{Schwarz88}-\cite{Tsubota00}. \

In spite of the recognized importance of the reconnecting loops kinetics,
the numerical results remain main source of information about this process.
The obvious lack of theoretical investigations interferes with deep insight
in the nature of this phenomena (this question had been recently discussed
in \cite{Barenghi2004}). For instance it is not clear how the cascade of
length in space of vortex loops sizes is formed, what mechanisms are
responsible for this, what quantities determine an intensity of cascade, and
why at all  breakdown of the loop prevails.

Of course the scarcity of analytic investigations related to
incredible complexity of the problem. Indeed we have to deal the
set of objects with not fixed number of elements, they can born
and die. Thus, some analog of the secondary quantization method is
required with the difference that objects (vortex loops)
themselves possess an infinite number of degree of freedom with
very involved dynamics. Clearly this problem can hardly be
resolved in nearest future. Recently in \cite {Copeland98} much
more modest approach, based on the ''rate equation'' for
distribution function $n(l)$ was elaborated in context of cosmic
strings. Following this work we introduce distribution function
$n(l,t)$ of density of loop in ''space'' of their lengths. It is
defined as a number of loops (per unit of volume) with lengths
between $l$ and $l+dl$. Due to reconnection processes $n(l,t)$ can
vary.

\begin{figure}[tbp]
\includegraphics[width=8cm]{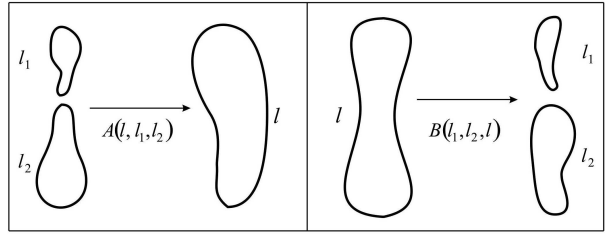}
\caption{Schematic sketch of the fusion and breaking-down of vortex loops.
Rates of these processes characterized by the coefficients $A(l_{1},l_{2},l)$
and $B(l,l_{1},l_{2})$.}
\label{A_B_loops}
\end{figure}

We discriminate two types of processes, namely the fusion of two
loops into the larger single loop  and the breakdown of  single
loop into two daughter loops. The kinetic of vortex tangle is
affected by the intensity of the introduced processes.
The intensity of the first process is characterized by the rate of collision $%
A(l_{1},l_{2},l)$ of two loops with lengths $l_{1}$and $l_{2}$ and forming
the loop of length $\ l=l_{1}+l_{2}$. The intensity of the second process is
characterized by the rate of self-intersection $B(l,l_{1},l_{2})$ of loop of
the length $l$ into two daughter loops with lengths $\ l_{1}$ and $l_{2}$
(see Fig.~\ref{A_B_loops}). In view of exposed above we can directly write
out the master ''kinetic'' equation for rate of change distribution function
$n(l,t)$.
\begin{align}
\frac{\partial n(l,t)}{\partial t}& =  \label{kinetic equation} \\
\int \int A(l_{1},l_{2},l)n(l_{1})n(l_{2})\delta
(l-l_{1}-l_{2})dl_{1}dl_{2}\;\;\;\;\;\;\;\;\;\;\;\;l_{1}+l_{2}& \rightarrow l
\notag \\
-\int \int A(l_{1},l,l_{2},)\delta
(l_{2}-l_{1}-l)n(l)n(l_{1})dl_{1}dl_{2}\;\;\;\;\;\;\;\;\;\;\;\;\;l_{1}+l&
\rightarrow l_{2}  \notag \\
-\int \int A(l_{2},l,l_{1},)\delta
(l_{1}-l_{2}-l)n(l)n(l_{1})dl_{1}dl_{2}\;\;\;\;\;\;\;\;\;\;\;\;l_{2}+l&
\rightarrow l_{1}  \notag \\
-\int \int B(l_{1},l_{2},l)n(l)\delta
(l-l_{1}-l_{2})dl_{1}dl_{2}\;\;\;\;\;\;\;\;\;\;\;\;\ \;\;\;\;\;\;\;\;\;l&
\rightarrow l_{1}+l_{2}\;\;\;\;\;  \notag \\
+\int \int B(l,l_{2},l_{1})\delta
(l_{1}-l-l_{2})n(l_{1})dl_{1}dl_{2}\;\;\;\;\;\;\;\
\;\;\;\;\;\;\;\;\;\;\;l_{1}& \rightarrow l+l_{2}  \notag \\
+\int \int B(l,l_{1},l_{2})\delta
(l_{2}-l-l_{1})n(l_{1})dl_{1}dl_{2}\;\;\;\;\
\;\;\;\;\;\;\;\;\;\;\;\;\;\;l_{2}& \rightarrow l+l_{1}  \notag
\end{align}
All of the processes are depicted at the left of each line. Clear that the
''kinetic'' equation has ''bookkeeping'' character. Physics of this approach
lies in the ''correct'' determinations of coefficient of that equations $%
A(l_{1},l_{2},l)$ and $B(l,l_{1},l_{2})$ on the base of some more
or less plausible model. \newline In the present work we derive
some general (independent on the model) relations for reconnection
rates $A(l_{1},l_{2},l)$ and $B(l,l_{1},l_{2}))$ ( Section II).
Then in Sec. III we apply these relations to evaluate the
reconnection rates on the base of Gaussian model of the vortex
tangle elaborated earlier \cite{Nemirovskii_97_1}. Section IV  is
devoted to conclusion and plans for future investigations.
\section{Mathematical identities for $A(l_{1},l_{2},l)$ and $%
B(l,l_{1},l_{2}).$} 
\begin{figure}[tbp]
\includegraphics[width=6cm]{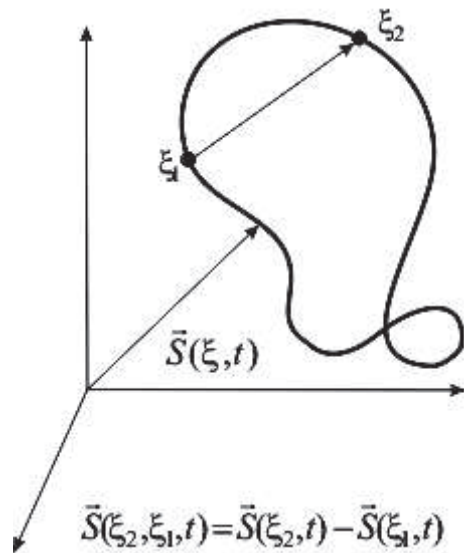}
\caption{Schematic sketch of vortex loop. Elements of line are described as
vectors $\mathbf{s}(\protect\xi )$, where the label variable $\protect\xi $
is taken here as the arc length. We associate the moment of intersection with the
vanishing of vector $\mathbf{S}(\protect\xi _{1},\protect\xi _{1},t)$
connecting points $\mathbf{s}(\protect\xi _{2},t)$ and $\mathbf{s}(\protect%
\xi _{1},t)$. \ }
\label{intersection}
\end{figure}

\begin{figure}[tbp]
\includegraphics[width=7cm]{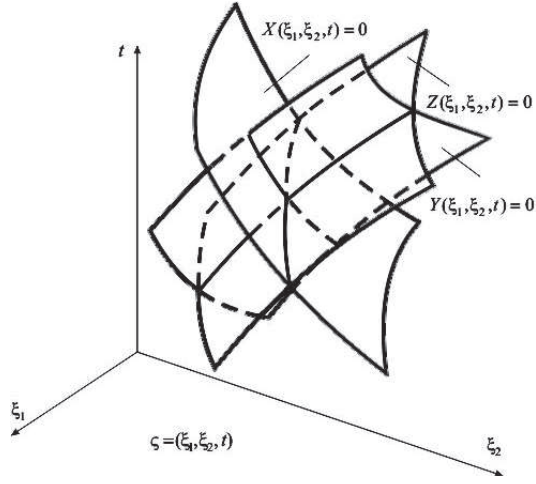}
\caption{The reconnection events can be described as zeroes of function $%
\mathbf{S}_{b}(\protect\xi _{2},\protect\xi _{1},t)$. In space of
its variables $\mathbf{\protect\zeta =\{}\protect\xi
_{2},\protect\xi _{1},t\}$ these points are cross of $3$ surfaces
${S}_{i}(\protect\xi _{2},\protect\xi _{1},t)=0,\;\;i=x,y,z$.}
\label{3_sufaces}
\end{figure}

In this section we will formulate mathematical
definition for quantities $\ A(l_{1},l_{2},l)$ and$\,$\
$B(l,l_{1},l_{2})$. We will start with the latter quantity
$B(l,l_{1},l_{2}).$ By
definition its physical meaning is just frequency of events when part of line with total length $%
l $ intersects the same line to create two two daughter loops with lengths $%
l_{1}$ and $l_{2}$ (see Fig.~\ref{A_B_loops}), that is
self-crossing event. As it is already stated we assume that each
crossing event leads to reconnection of lines. The elements of
vortex line are described as a function $\mathbf{s}(\xi,t)$ \ (See
Fig.~\ref{intersection}) that is time dependant radius-vector of
the points resting on loop. Variable $\xi $ labels the points of
the loop. It is convenient to choose variable $\xi $ to be equal
to the arc length, $(0\leq \xi \leq l)$. Let us consider function
\begin{equation}
\mathbf{S}_{s}(\xi _{2},\xi _{1},t)=\mathbf{s}(\xi _{2},t)-\mathbf{s}(\xi
_{1},t),  \label{distance_single}
\end{equation}
that is just $3D$ vector connecting points $\mathbf{s}(\xi
_{2},t_{2})$ and $\mathbf{s}(\xi _{1},t_{1})$. Clearly that
condition $\mathbf{S}_{b}(\xi _{2},\xi _{1},t)=0$ implies that the
self-crossing event of parts of line with label-coordinates $\xi
_{2},\xi _{1}$ occurs at moment of time $t$. \ The quantity
$\mathbf{S}_{b}(\xi _{2},\xi _{1},t)$ is fluctuating 3-component
function of three arguments $\xi _{2},\xi _{1},t$. We are
interested in how often $\mathbf{S}_{b}(\xi _{2},\xi _{1},t)$
vanishes
in cube of space $\mathbf{\zeta =\{}\xi _{2},\xi _{1},t\}$. Points in space $%
\mathbf{\zeta =\{}\xi _{2},\xi _{1},t\}$ where function
$\mathbf{S}_{b}(\xi _{2},\xi _{1},t)$ vanishes, are points of
cross of $3$ surfaces ${S}_{i}(\xi _{2},\xi _{1},t)=0,\;\;i=x,y,z$
in space $\mathbf{\zeta =\{}\xi _{2},\xi _{1},t\}$ as it is shown
in Fig.~\ref{3_sufaces}.

From theory of generalized function it follows that number of these points
(we denote them below as $\mathbf{\zeta }_{a}$) can be expressed via $\delta
$-function of quantity $\mathbf{S}_{b}(\xi _{2},\xi _{1},t\dot{)}$ with the
help of following formula .
\begin{equation}
\sum_{a} \delta (\mathbf{\zeta }-\mathbf{\zeta }_{a})=\left| \frac{\partial
(X,Y,Z)}{\partial (\xi _{2},\xi _{1},t)}\right| _{\mathbf{\zeta }=\mathbf{%
\zeta }_{a}}\delta (\mathbf{S}_{b}(\xi _{2},\xi _{1},t))
\label{s_zeroes_number}
\end{equation}
Here $X,Y,Z$ are the components of vector $\mathbf{S}_{b}(\xi _{2},t_{2},\xi
_{1},0) $ . By integration of both parts of (\ref{s_zeroes_number}) over $%
d\xi _{1}d\xi _{2}$ we would obtain the full number of
intersections (per unit time). If, further to introduces
additional constraint $\delta (\xi _{2}-\xi _{1}-l_{1})\,$\
implying that the distance (along line) between chosen points is
equal to $l$ and to integrate over $d\xi _{1}d\xi _{2}$, we obtain
the rate of self-intersection of line with length $l$ and
breakdown it into pieces $l_{1}$ and $l-l_{1}$.

In addition we have to do averaging \ over all possible fluctuating
configurations. Thus the coefficient $B(l_{1},l-l_{1},l)$ with dimension $%
\left[ s\right] =s^{-1}cm^{-1}$ is equal to
\begin{equation}
B(l_{1},l-l_{1},l)=\int \int d\xi _{1}d\xi _{2}\delta (\xi _{2}-\xi
_{1}-l_{1})\,\left\langle \left| \frac{\partial (X,Y,Z)}{\partial (\xi
_{2},\xi _{1},t)}\right| _{\mathbf{\zeta }=\mathbf{\zeta }_{a}}\delta (%
\mathbf{S}_{b}(\xi _{2},\xi _{1},t))\right\rangle .  \label{B_def}
\end{equation}

To obtain coefficient $A(l_{1},l_{2},l)$ we use the similar procedure. Let
us consider two loops with length $l_{1}$ and $l_{2}$. Our purpose now to
find the rate $A(l_{1},l_{2},l)$ of fusion of these two loops into one loop
of length $l=l_{1}+l_{2}$. Dimension of $A(l_{1},l_{2},l)$ is $%
[A]=cm^{3}s^{-1}$. As previously we describe vortex filaments by positions
of radius vectors their elements $\mathbf{s}(\xi _{1},t)$ and $\mathbf{s}%
(\xi _{2},t)$. Here we have two label variables $\xi _{1},\xi _{2}$
belonging to different loops and running in limits $(0\leq \xi \leq l_{1})$
and $(0\leq \xi \leq l_{2})$ respectively. One more important difference
with the previous case is that both functions $\mathbf{s}(\xi _{1},t)$ and $%
\mathbf{s}(\xi _{2},t)$ should depend on ''initial'' positions $\mathbf{s}%
(\xi _{2}=0,t)=\mathbf{R}_{1}(t)$ and $\mathbf{s}(\xi _{2}=0,t)=\mathbf{R}%
_{2}(t)$, chosen arbitrary. Of course in previous case of
self-intersection of single loop, quantity $\mathbf{s}(\xi ,t)$
also depended on ''initial'' positions $\mathbf{R(t)}$, but it did
not influence the rate of self-intersection. Now for case of the
fusion this dependance is important, since very distant loops have
small probability to collide. Let us introduce the ''fusion''
functions
\begin{equation}
\mathbf{S}_{f}(\xi _{2},\xi _{1},t_{1})=\mathbf{s}(\xi _{2},t)-\mathbf{s}%
(\xi _{1},t).  \label{distance_double}
\end{equation}
Repeating the considerations for case of the single loop we find that number
of reconnection (per unit of time) of points $\xi _{2},\xi _{1}$ formally
coincides with (\ref{s_zeroes_number})\bigskip
\begin{equation}
\sum \delta (\mathbf{\zeta }-\mathbf{\zeta }_{a})=\left| \frac{\partial
(X,Y,Z)}{\partial (\xi _{2},\xi _{1},t)}\right| _{\mathbf{\zeta }=\mathbf{%
\zeta }_{a}}\delta (\mathbf{S}_{f}(\xi _{2},\xi _{1},t))
\label{m_zeroes_number}
\end{equation}
with the difference that $\xi _{2},\xi _{1}$ belong to different curves.
Since intersections of any elements of lines lead to fusion of the loops we
have to integrate (\ref{m_zeroes_number}) over $d\xi _{1}d\xi _{2}$. The
result obtained is valid for chosen pair of loops. To obtain the
total number of events we have to multiply the result obtained by quantity $ n(l_{1})n(l_{2})d
\mathbf{R}_{1}d\mathbf{R}_{2}$, which is just a full number of loops of chosen sizes in the whole volume.
Comparing with the master kinetic equation (%
\ref{kinetic equation}) we find the final expression for fusion coefficient $%
A(l_{1},l_{2},l)$

\begin{equation}
A(l_{1},l_{2},l)=\frac{1}{\text{VOLUME}}\int \int d\mathbf{R}_{1}d\mathbf{R}%
_{2}\int \int d\xi _{1}d\xi _{2}\,\left\langle \left| \frac{\partial (X,Y,Z)%
}{\partial (\xi _{2},\xi _{1},t)}\right| _{\mathbf{\zeta }=\mathbf{\zeta }%
_{a}}\delta (\mathbf{S}(\xi _{2},\xi _{1},t))\right\rangle .  \label{A_def}
\end{equation}

Thus we obtained expressions (\ref{B_def})and (\ref{A_def}), which
allow to calculate the rates of reconnections of the fusion and
breakdown vortex loops. They are however just formal mathematical
identities. Concrete results depend on statistics of individual
lines. Therefore to move further we have to ascertain the
procedure for averaging. It will be done in the next section.

\section{Gaussian model case}

\subsection{\protect\bigskip The generalized Wiener distribution \ }

\begin{figure}[tbp]
\includegraphics[width=7cm]{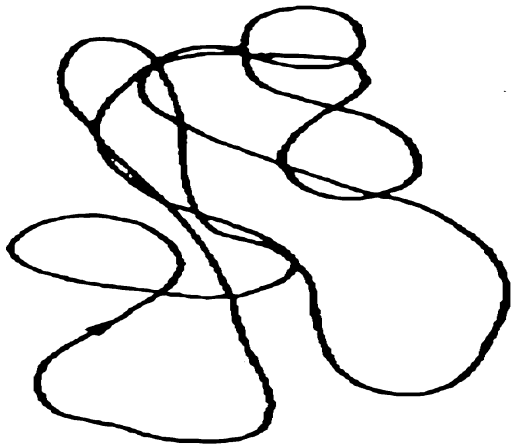}
\caption{Snapshot of the ''average'' vortex loop obtained from analysis of
the statistical properties. Close $(\Delta \protect\xi \ll R)$ parts of the
line are separated in $3D$ space by distance $\Delta \protect\xi .$ The
distant parts $(R\ll \Delta \protect\xi )$ are separated in $3D$ space by
the distance $\protect\sqrt{\protect\xi _{0}\Delta \protect\xi }$, i.e. the
vortex loop has the typical random walking structure. }
\label{RW}
\end{figure}

To evaluate quantities $B(l_{1},l_{2},l)$ and $A(l_{1},l_{2},l)$ written in
form (\ref{B_def}) and (\ref{A_def}) one needs to know statistics of
individual loops. In general case this statistics should be extracted from
investigation of the full dynamical problem. The according statement of such
problem includes equation of motion (Biot-Savart law for quantum vortices),
dissipative effects (interaction with normal component) and additional
Langevin force responsible for chaotic behavior. The problem becomes more
involved by circumstance that other loops also influence dynamics of the
chosen loop. At this stage we choose another way, namely we use the Gaussian
model of the vortex tangle elaborated by author \cite{Nemirovskii_97_1}. To
develop this approach the trial distribution functional of Gaussian form had
been consructed. This functional absorbed all  properties of the superfluid
turbulence known from both experimental studies and numerical simulations.
According to this model the ''average'' vortex loop has a
typical structure shown in Fig.~\ref{RW}.
The close parts of the loop separated (along line) by distance
$\xi _{2}-\xi _{1}$ smaller then the mean radius of curvature $\xi
_{0}$ are strongly correlated, $\left\langle \mathbf{s}^{\prime
}(\xi _{1},t)\mathbf{s}^{\prime }(\xi _{2},t)\right\rangle
\rightarrow 1,$ ($\mathbf{s}^{\prime }$ is the tangent vector) and
line is smooth. \ Remote parts of the line $\ \ (\xi _{2}-\xi
_{1}\gg \xi _{0})$ are not correlated at all, $\left\langle \mathbf{s}%
^{\prime }(\xi _{1},t)\mathbf{s}^{\prime }(\xi _{2},t)\right\rangle
\rightarrow 0$. Thus for large separations the vortex loop has a typical
''random walking structure'' with the Wiener distribution. This
''semifractal'' behavior satisfies to the generalized \ Wiener distribution.
Namely, the probability $\mathcal{P}(\left\{ \mathbf{s}(\xi ,t)\right\} )$\
to find some particular configuration $\left\{ \mathbf{s}(\xi ,t)\right\} $
is expressed by the probability distribution functional (see for details in
\cite{Nemirovskii_97_1})
\begin{equation}
\mathcal{P}(\left\{ \mathbf{s}(\xi ,t)\right\} )=\mathcal{N}\exp
\left( -\int\limits_{0}^{l}\int\limits_{0}^{l}\mathbf{s}^{\prime
\alpha }(\xi _{1},t)\Lambda ^{\alpha \beta }(\xi _{1}-\xi
_{2})\mathbf{s}^{\prime \beta }(\xi _{2},t)d\xi _{1}d\xi
_{2}\right) .  \label{Gauss_model}
\end{equation}
Here $\mathcal{N}$ is normalizing factor, $l$ is the length of
curve, it is supposed to be much larger then mean radius of
curvature $\xi _{0}$. The probability distribution functional \
(\ref{Gauss_model}) introduced in \cite {Nemirovskii_97_1}
distribution considered  anisotropy and \ polarization of vortex
tangle in counterflowing turbulent helium. This fact is reflected
by the circumstance that $\Lambda ^{\alpha \beta }(\xi _{1}-\xi
_{2})$ is the full matrix. In practice it is more convenient to
deal with characteristic functional $W(\{\mathbf{P}(\xi,t)\})$
defined as
\begin{equation}
W(\{\mathbf{P}(\xi,t)\})\;=\left\langle \exp \left( i\int\limits_{0}^{l}%
\mathbf{P}(\xi ,t)\mathbf{s}^{\prime }(\xi ,t)d\xi \ \right) \right\rangle
\;.  \label{CF_def}
\end{equation}
The characteristic functional enables us calculate any averages depending on
vortex lines configuration $\left\{ \mathbf{s}(\xi ,t)\right\} $ by simple
functional differentiation. For instance the average tangent vector $%
\left\langle \mathbf{s}_{\alpha }^{\prime }(\xi _{1})\right\rangle $ or the
correlation function between orientation of the different elements of the
vortex filaments $\left\langle \mathbf{s}_{\alpha }^{\prime }(\xi _{1})%
\mathbf{s}_{\beta }^{\prime }(\xi _{2})\right\rangle $ are readily expressed
via characteristic functional accordingly to the following rules:

\begin{equation}
\left\langle \mathbf{s}_{\alpha }^{\prime }(\xi _{1})\right\rangle =\;\left.
\frac{\delta W}{i\delta \mathbf{P}^{\alpha }(\xi _{1})}\right| _{\;\mathbf{P}%
\;=\;0},\;\;\;\;\;\left\langle \mathbf{s}_{\alpha }^{\prime }(\xi _{1})%
\mathbf{s}_{\beta }^{\prime }(\xi _{2})\right\rangle =\;\left. \frac{\delta
^{2}W}{i\delta \mathbf{P}^{\alpha }(\xi _{1})\;\;i\delta \mathbf{P}^{\beta
}(\xi _{2})}\right| _{\mathbf{P}\;=\;0}  \label{r'(W)}
\end{equation}
Calculation of the characteristic functional $W(\{\mathbf{P}(\xi ,t)\})$  (%
\ref{CF_def}) on the base of the probability functional \
(\ref{Gauss_model}) is reduced to functional integration, which,
in turn, reduces to the ''full square procedure''. \ The result is

\begin{equation}
W(\{\mathbf{P}(\xi,t)\})\;=\exp \left\{
-\int\limits_{0}^{l}\int\limits_{0}^{l}\mathbf{P}^{\alpha }(\xi
_{1})N^{\alpha \beta }(\xi _{1}-\xi _{2})\mathbf{P}^{\beta }(\xi _{2})d\xi
_{1}d\xi _{2}\right\}.  \label{CF_N}
\end{equation}

To avoid unnecessary lengthy calculations we simplify the model
expressed by the probability distribution functional \
(\ref{Gauss_model}), namely we omit both the anisotropy and \
polarization. In this case the matrix $N^{\alpha \beta }(\xi
_{1}-\xi _{2})$ used in (\cite{Nemirovskii_97_1}) can be taken as
\begin{equation}
N^{\alpha \beta }(\xi _{1}-\xi _{2})=\delta _{\alpha \beta }\frac{1}{6\left(
1-2\xi _{0}\sqrt{\pi }/l\right) }\left( \exp \left[ -\frac{(\xi _{1}-\xi
_{2})^{2}}{4\xi _{0}^{2}}\right] -\frac{2\xi _{0}\sqrt{\pi }}{l}\right) .
\label{N_isotrop}
\end{equation}
For small separation $\ (\xi _{2}-\xi _{1}\ll \xi _{0})$ , $%
\sum\nolimits_{\alpha }N^{\alpha }(\xi _{1}-\xi _{2})\rightarrow
1/2$ \ that guarantees that $\left\langle \mathbf{s}^{\prime }(\xi
,t)\mathbf{s}^{\prime }(\xi,t)\right\rangle =1,$ as it should be
for smooth lines (we recall that for pure random walking \ line
the quantity $\left\langle \mathbf{s}^{\prime }(\xi
,t)\mathbf{s}^{\prime }(\xi ,t)\right\rangle $ does not exist at
all). For large separation $(\xi _{2}-\xi _{1}\gg \xi _{0})$
exponents tends to $\delta _{\alpha \beta }(2\sqrt{\pi } \xi
_{0}/6)\delta (\xi _{1}-\xi _{2}),$ and correlation between
tangent
vectors weakens,\ $\left\langle \mathbf{s}^{\prime }(\xi ,t)\mathbf{s}%
^{\prime }(\xi ,t)\right\rangle \rightarrow 0.$ \ The second term
in parenthesis of expression (\ref{N_isotrop}) appears due to
closeness of the loops. It ensures that
\begin{equation*}
\int\limits_{0}^{l}\int\limits_{0}^{l}d\xi _{1}d\xi _{2}\left\langle \mathbf{%
s}^{\prime }(\xi _{1},t)\mathbf{s}^{\prime }(\xi _{2},t)\right\rangle
=\left\langle (\mathbf{s}(l,t)-\mathbf{s}(0,t))^{2}\right\rangle =0
\end{equation*}
as it should be for the closed lines (see explanations in (\cite
{Nemirovskii_97_1})). Thus the characteristic functional with
function $\ N$ satisfies to necessary ''semifractal'' behavior of
line \ and will be used further for evaluation of the rate
reconnection coefficients. Here it is necessary to do two
important comments. First one concerns the ''semifractal''
structure of the vortex loop. In principle, reconnections events
occur due random walking behavior with non-zero probability of the
meeting of any two elements. However, to evaluate the according
rates on base of relations (\ref{B_def}), (\ref{A_def}), we need
to know not only an average of $\delta (\mathbf{S}(\xi _{2},\xi
_{1},t))$ but also an average of the Jacobian. But the latter
includes averaged derivatives (with respect two both label
variable$\xi _{2},\xi _{1}$ and time). It is known fact however
that quantities like $\left\langle \left| \mathbf{s}_{\alpha
}^{\prime }(\xi _{1})\right| \right\rangle ,\;\left\langle
\mathbf{s}_{\alpha }^{\prime }(\xi _{1})\mathbf{s}_{\beta
}^{\prime }(\xi _{2})\right\rangle $ etc. are ill-defined for pure
random walking structure, For that reason we need to smooth away
the curve on small scales. The second remark concerns the mean
radius of curvature $\xi _{0}$. This quantity\ appears here as
important parameter of theory of random walk. It is frequently \
called as ''an elementary step'' or persistence length. In fact
the theory described by the Wiener distribution is failed for
small scales less then $\xi _{0}$ , therefore usually this value
appears as a low cut-off of the whole approach.

\section{\protect\bigskip Evaluation of $A(l_{1},l_{2},l)$ and $%
B(l,l_{1},l_{2}).$}

\subsection{evaluation of $\ B(l,l_{1},l_{2})$}

Let us come back to the rate coefficients. We start with the
self-intersection processes. Positions of line elements $\mathbf{s}(\xi
_{2},t),\mathbf{s}(\xi _{1},t)$ and relative vector $\mathbf{S}_{b}(\xi
_{2},\xi _{1},t)$ are strongly fluctuating quantities having the Gaussian
statistics. Due to the Wick theorem \ the average in integrand of (\ref
{B_def}) \ can be taken as a sum of all possible pairs of quantity $\mathbf{S%
}_{b}(\xi _{2},t,\xi _{1},t)\,\ \ $and its derivatives. Because of
uniformity in $\xi $ space, quantity $\mathbf{S}_{b}(\xi _{2},\xi _{1},t)\,$%
\ depends on $\left| \xi _{2}-\xi _{1}\right| ,$ for that reason all
averages of structure $\left\langle (\partial X/\partial \xi _{1})\delta
(X(\xi _{2},\xi _{1},t))\right\rangle $ vanish, therefore only pairs
separately from $\mathbf{S}_{b}(\xi _{2},\xi _{1},t)\,$ and from its
derivatives survive. As a result the average of production is equal to
production of \ averages and each of the factors can be evaluated separately
\begin{equation}
\left\langle \left| \frac{\partial (X,Y,Z)}{\partial (\xi _{2},\xi _{1},t)}%
\right| _{\mathbf{\zeta }=\mathbf{\zeta }_{a}}\delta (\mathbf{S}_{b}(\xi
_{2},\xi _{1},t))\right\rangle =\,\left\langle \left| \frac{\partial (X,Y,Z)%
}{\partial (\xi _{2},\xi _{1},t)}\right| _{\mathbf{\zeta }=\mathbf{\zeta }%
_{a}}\right\rangle \left\langle \delta (\mathbf{S}_{b}(\xi _{2},\xi
_{1},t))\right\rangle.  \label{self_separate}
\end{equation}
As mentioned, the use of the characteristics functional (\ref{CF_def}), (\ref
{CF_N}) allows to calculate any averaged functional of configurations $%
\left\{ \mathbf{s}(\xi ,t)\right\}.$ Let us show  how to evaluate $%
\left\langle \delta \mathbf{S}_{s}(\xi _{2},\xi _{1},t)\right\rangle $. With
use of the standard integral representation for $\delta $-function
\begin{equation*}
\delta (x)=\frac{1}{(2\pi )}\int_{-\infty }^{\infty }e^{ixy}dy,
\end{equation*}
\bigskip we rewrite $\left\langle \delta \mathbf{S}_{b}(\xi _{2},t,\xi
_{1},t)\right\rangle $ as
\begin{eqnarray}
\left\langle \delta \mathbf{S}_{b}(\xi _{2},\xi _{1},t)\right\rangle &=&%
\frac{1}{(2\pi )^{3}}\int \left\langle \exp \left[ i\mathbf{y(s(}\xi
_{2},t_{2}\mathbf{)-s(}\xi _{1},t_{1}\mathbf{))}\right] \right\rangle d^{3}%
\mathbf{y=}  \notag \\
&&\frac{1}{(2\pi )^{3}}\int \left\langle \exp \left( i\int\limits_{\xi
_{1}}^{\xi _{2}}\mathbf{ys}^{\prime }(\xi ,t)d\xi dt\ \right) \right\rangle
d^{3}\mathbf{y.}  \label{Delta_integral}
\end{eqnarray}
Comparing (\ref{CF_def}) and (\ref{Delta_integral}) we conclude
that the integrand in last term of (\ref{Delta_integral}) is just
the characteristic functional $W(\{\mathbf{P}(\xi ,t)\}),$ taken
at value of $\mathbf{P}(\xi ,t) $
\begin{equation}
\mathbf{P}(\xi )\;=\;-\mathbf{y}\theta (\xi -\xi _{1})\theta (\xi _{2}-\xi ).
\label{TETA}
\end{equation}
Here $\theta (\xi )$ is the unit step-wise function. Relation (\ref{TETA})
implies that we choose in integrand of the characteristic functional only
points  lying in interval from $\xi _{1}$ to $\xi _{2}$ on \ the curve.
Substituting this value $\mathbf{P}(\xi ,t)$ expressed by (\ref{TETA}) in CF
(\ref{CF_N}) with function \ $N^{\alpha \beta }(\xi _{1}-\xi _{2})$ in form (%
\ref{N_isotrop}) we obtain
\begin{equation}
\left\langle \delta \mathbf{S}_{b}(\xi _{2},t,\xi _{1},t)\right\rangle =%
\frac{1}{(2\pi )^{3}}\int \exp \left[ -\mathbf{y}^{2}\frac{2\sqrt{\pi }\xi
_{0}}{6}\left( (\xi _{2}-\xi _{1})-(\xi _{2}-\xi _{1})^{2}/l\right) \right]
d^{3}\mathbf{y=}\left( \frac{3}{2\pi ^{3/4}\xi _{0}\left( (\xi _{2}-\xi
_{1})-(\xi _{2}-\xi _{1})^{2}/l\right) }\right) ^{3/2}  \label{delta_self}
\end{equation}
Evaluation of absolute value of Jacobian in (\ref{self_separate}) we perform
by use of $\left| J\right| =\sqrt{J^{2}}.$ Furthermore it would be
convenient for the sake of generalization to use vector velocity $\mathbf{V}%
_{l}=d\mathbf{s}/dt,$ instead of calculation of it in explicit form,
expressing velocity via the vortex filament configuration $\{\mathbf{s}(\xi )\}.$
Calculation of $J^{2}$ can be fulfilled writing \ Jacobian in explicit form
and subsequent applying the Wick theorem. Simple but tedious calculations
leads to result that
\begin{equation}
J^{2}=2\left\langle \mathbf{V}_{lx}^{2}\right\rangle \left\langle (\partial
\mathbf{s}_{y}\mathbf{/}\partial \xi _{1})^{2}\right\rangle \left\langle
(\partial \mathbf{s}_{z}\mathbf{/}\partial \xi _{2})^{2}\right\rangle
+2\left\langle \mathbf{V}_{lx}^{2}\right\rangle \left\langle (\partial
\mathbf{s}_{y}\mathbf{/}\partial \xi _{2})^{2}\right\rangle \left\langle
(\partial \mathbf{s}_{z}\mathbf{/}\partial \xi _{1})^{2}\right\rangle +p.p.
\label{J_squared}
\end{equation}
where $p.p.$ all permutations with respect to $x,y,z$. \ Estimating $%
\left\langle (\partial \mathbf{s}_{y}\mathbf{/}\partial \xi
_{1})^{2}\right\rangle $ and similar terms as $1/3$ we obtain that $\left|
J\right| =\frac{2}{3}\left| \mathbf{V}_{l}\right| .$\bigskip\ After use of
the integration $\int \int d\xi _{1}d\xi _{2}\delta (\xi _{2}-\xi
_{1}-l_{1})\,$\ (see \ref{B_def}) we finally obtain
\begin{equation}
B(l_{1},l-l_{1},l)=b_{s}\ast V_{l}\frac{l}{(\xi
_{0}(l_{1}-l_{1}^{2}/l))^{3/2}}  \label{B_final}
\end{equation}
where constant $b_{s}=\left( \sqrt{3}/8\pi ^{9/4}\right) \approx
0.0164772.$ We introduced in coefficient $B$ the additional factor
$1/2$ in order to avoid the over-counting of the reconnection
events, since decays $l\rightarrow l_{1}+l_{2}$ and $l\rightarrow
l_{2}+l_{1}$ describe the same process, though the both enter into
equations. Let us recall that the quantity $\xi _{0}$ is the mean
radius of curvature. The numerical factor $b_{s}\approx 0.0164772$
is the result of particular approximation used in paper (\cite
{Nemirovskii_97_1}). If for instance we used pure Wiener
distribution \ with the persistency length $\ \xi _{0}$ and
disregard the closeness of line,  we would have for coefficient
$B(l_{1},l-l_{1},l)$ the expression
\begin{equation}
B_{C}(l_{1},l-l_{1},l)=b_{s}\ast V_{l}\frac{l}{(\xi _{0}l_{1})^{3/2}}
\end{equation}
where $b_{s}\approx 0.11.$ This result is very similar to result
obtained earlier in paper \cite{Copeland98} from the qualitative
consideration. Yet it is important also that in this way it is not
clear how to relate the persistency length $\ \xi _{0}$ with the
mean radius of curvature.

\subsection{\protect\bigskip evaluation of $A(l_{1},l_{2},l)$}

Let us now evaluate quantity $A(l_{1},l_{2},l)$ defined by
relation (\ref {A_def}). We again (as for the previous case)
evaluate average from Jacobian and $\delta $-function separately.
Contribution from Jacobian coincides with the previous result\
$\left| J\right| =\frac{2}{3}\left| \mathbf{V}_{l}\right| $. The
rest $\delta $-function part can be evaluated with the help of the
CF obtained above. Unlike previous case we have to know two-loop
distribution function. Since we omit interaction of loops (until
the reconnection event occurs) the CF \ for two loops with lengths $l_{1}$%
and $\ l_{2}$ is just production of the expressions of type (\ref{CF_N})

\begin{eqnarray}
W(\{\mathbf{P}_{1}(\xi )\},\{\mathbf{P}_{2}(\xi )\})\; &=&\exp \left\{
-\int\limits_{0}^{l_{1}}\int\limits_{0}^{l_{1}}\mathbf{P}^{\alpha }(\xi
_{1})N_{1}^{\alpha \beta }(\xi _{1}-\xi _{2})\mathbf{P}^{\beta }(\xi
_{2})d\xi _{1}d\xi _{2}\right\} \times  \label{W_double} \\
&&\exp \left\{ -\int\limits_{0}^{l_{2}}\int\limits_{0}^{l_{2}}\mathbf{P}%
^{\alpha }(\xi _{1})N_{2}^{\alpha \beta }(\xi _{1}-\xi _{2})\mathbf{P}%
^{\beta }(\xi _{2})d\xi _{1}d\xi _{2}\right\} .  \notag
\end{eqnarray}
Quantities $N_{1}^{\alpha \beta }(\xi _{1}-\xi _{2})$ and $N_{2}^{\alpha
\beta }(\xi _{1}-\xi _{2})$ differ from each other only by lengths of loops $%
l_{1}$and $\ l_{2}$, entering expressions for $N_{1}^{\alpha \beta
}$.
Further, by use of the standard integral representation for $\delta $%
-function we have \bigskip
\begin{equation}
\left\langle \delta \mathbf{S}_{f}(\xi _{2},\xi _{1},t)\right\rangle =\frac{1%
}{(2\pi )^{3}}\int \left\langle \exp \left[ i\mathbf{y(s}_{2}\mathbf{(}\xi
_{2},t_{2}\mathbf{)-s}_{1}\mathbf{(}\xi _{1},t_{1}\mathbf{))}\right]
\right\rangle d^{3}\mathbf{y.}  \label{delta_fusion}
\end{equation}
We stress again that the label variables $\xi _{2}$ and $\xi _{1}$ belongs
two different loops. Let us introduce initial points $\mathbf{s}_{1}\mathbf{(%
}0\mathbf{)}$ and $\mathbf{s}_{2}\mathbf{(}0\mathbf{)}$ and rewrite (\ref
{delta_fusion}) in the following form:
\begin{equation*}
\frac{1}{(2\pi )^{3}}\int \exp \left[ -i\mathbf{y(s}_{2}\mathbf{(}0\mathbf{%
)-s}_{1}\mathbf{(}0\mathbf{))}\right] \left\langle \exp \left[ i\mathbf{y(s}%
_{2}\mathbf{(}\xi _{2}\mathbf{)-s}_{2}\mathbf{(}0\mathbf{))}\right] \exp %
\left[ -i\mathbf{y(s}_{1}\mathbf{(}\xi _{2}\mathbf{)-s}_{1}\mathbf{(}0%
\mathbf{))}\right] \right\rangle d^{3}\mathbf{y.}
\end{equation*}
Identifying further \ the ''initial'' positions $\mathbf{s}_{1}\mathbf{(}0%
\mathbf{),s}_{2}\mathbf{(}0\mathbf{)}$ with quantities $\mathbf{R}_{1},%
\mathbf{R}_{2}$ in formula (\ref{A_def}) we rewrite it as
\begin{eqnarray}
A(l_{1},l_{2},l) &=&\frac{1}{\text{VOLUME}}\frac{2}{3}\left| \mathbf{V}%
_{l}\right| \int \int d\mathbf{R}_{1}d\mathbf{R}_{2}\int \int d\xi _{1}d\xi
_{2}  \label{A_inter} \\
&&\frac{1}{(2\pi )^{3}}\int \exp \left[ -i\mathbf{y(\mathbf{R}_{2}-\mathbf{R}%
_{1})}\right] \left\langle \exp \left[ i\mathbf{y(s}_{2}\mathbf{(}\xi _{2}%
\mathbf{)-s}_{2}\mathbf{(}0\mathbf{))}\right] \exp \left[ -i\mathbf{y(s}_{1}%
\mathbf{(}\xi _{2}\mathbf{)-s}_{1}\mathbf{(}0\mathbf{))}\right]
\right\rangle d^{3}\mathbf{y}.  \notag
\end{eqnarray}
Let us introduce variables $\mathbf{R}_{1}-\mathbf{R}_{2},(\mathbf{R}_{1}+%
\mathbf{R}_{2})/2.$ \ Integration over $\mathbf{\mathbf{R}}_{2}\mathbf{-%
\mathbf{R}}_{1}$ gives $\mathbf{\delta }\mathbf{(y)}$, integration over $(%
\mathbf{R}_{1}+\mathbf{R}_{2})/2$ gives the total volume of
system. Further, integration over $\mathbf{y}$ gives unity, and
integration over $\xi _{1},\xi _{2}$ gives the production
$l_{1}l_{2}$. \ Thus we obtain the remarkable result, that for
noninteracting loops quantity $A(l_{1},l_{2},l)$ \ responsible for
merging of loops does not depend on statistics of individual loop
at all and is equal
\begin{equation}
A(l_{1},l_{2},l)=b_{m}V_{l}l_{1}l_{2}  \label{A_final}
\end{equation}
Here $b_{m}=1/3.$ As earlier we introduced additional factor $1/2$ to avoid the over-counting of the reconnection events.

Results (\ref{A_final}) and (\ref{B_final}) \ (with not well determined
factors $b_{s}$ and $b_{m}$ ) were also obtained in paper \cite{Copeland98}.
Authors used some qualitative picture of moving and colliding elements of
lines. This fact confirms the validity of approach made in our work,
which allows to use it for more complicated (in comparison with random walk)
cases.

\section{Conclusion}

\bigskip In the paper we discussed a role of kinetics of the merging and
breaking down vortex loops in the whole vortex tangle dynamic. Following the
work \cite{Copeland98} we introduced the reconnection rates $%
A(l_{1},l_{2},l) $ and $B(l,l_{1},l_{2})$. Quantity $A$ is a number (per
unit of time and per \ unit of volume) of the fusion of two loops of lengths
$l_{1}$and $l_{2}$ with formation of the single loop of length $\
l=l_{1}+l_{2}$. Quantity $B(l,l_{1},l_{2})$ describes rate of the breakdown
of \ the loop with the length $l$ into two daughter loops with lengths $\
l_{1}$ and $l_{2}$. We developed mathematical formalism, which enables us to
calculate $A(l_{1},l_{2},l)$ and $B(l,l_{1},l_{2}).$ Briefly, these
quantities can be found as an average number of zeroes of vector $\mathbf{S}%
_{b}(\xi _{2},\xi _{1},t)$ connecting points of $\xi _{2}$ and $\ \xi _{1}$
at moment of time $t$. The averaging procedure should be fulfilled
by using statistics of individual loops, expressed e.g. by the probability
distribution functional $\mathcal{P}(\left\{ \mathbf{s}(\xi ,t)\right\} ).%
\mathcal{\ }$\ We applied this formalism to evaluate the reconnection rates $%
A(l_{1},l_{2},l)$ and $B(l,l_{1},l_{2})$ in case when vortex loop
is semifractal object, which is smooth for small separations along
the line and has a random work structure for distant (along line)
points.\\
 Relations (\ref{A_def}) and (\ref{B_def})are the key
results of the paper. They are formal, and do not depend on
statistic of loop. They can be applied for any physical situation
provided that we are able to get statistics from
the  dynamical problem. In that sense the application of the results (%
\ref{A_def}) and (\ref{B_def}) for Gaussian case is just
illustration how to use it in concrete sense. One of serious
problem in this illustration is that in Gaussian approach the low
scale was restricted by mean radius of curvature. In real vortex
tangle $\xi _{0}$ in superfluid turbulent HeII this scale exceeds
greatly the scales at \ which the real dissipative mechanisms
occur. Thus, the study of what would happen for scales smaller of
mean radius of curvature is on of actual problems. Nevertheless
results expressed by relations (\ref{A_final}) and (\ref{B_final})
are significant, they allows to analyze the kinetic equation
(\ref{kinetic equation}), the attempts were made in papers
\cite{Copeland98}, \cite{Sreer99}. We also will apply these result
in the forthcoming paper for study of the kinetic
equation.\newline

I am grateful to participants of the workshop ''Superfluidity
under Rotation'' (Manchester, 2005) for useful discussion of the
results exposed above. This work was partially supported by grant
N 03-02-16179 of the Russian Foundation of Fundamental Research.

\end{document}